 \documentclass [12pt,a4paper      ]{article}
\usepackage{graphics}
\usepackage{times}

\DeclareFontFamily{OT1}{times}{}
\DeclareFontShape {OT1}{times}{m }{n }{ <-> ptmr }{}
\DeclareFontShape {OT1}{times}{bx}{n }{ <-> ptmb }{}
\DeclareFontShape {OT1}{times}{m }{it}{ <-> ptmri}{}
\DeclareFontShape {OT1}{times}{bx}{it}{ <-> ptmbi}{}
\usepackage{amsmath}
\usepackage{amsfonts}
\usepackage{amssymb}
\usepackage{latexsym}
\numberwithin{equation}{section}

\begin{document}

\title{{\bf \vspace{-2.5cm} Antimatter weapons (1946-1986):\\ From Fermi and Teller's speculations to the first open scientific publications} \footnote{Expanded version of a paper published in French in La Recherche {\bf 17}  (Paris, 1986) 1440--1443; in English in The World Scientist (New Delhi, India, 1987) 74--77, and in Bulletin of Peace Proposals {\bf 19}  (Oslo,1988) 444--450; and in Finnish in Antimateria-aseet (Kanssainv\"alinen rauhantutkimuslaitos, Helsinki, 1990, ISBN 951-9193-22-7) 7--18.}}
%  =======================================================

\author{{\bf Andre Gsponer and Jean-Pierre Hurni}\\
\emph{Independent Scientific Research Institute}\\ 
\emph{Box 30, CH-1211 Geneva-12, Switzerland}\\
e-mail: isri@vtx.ch\\}

\date{Version ISRI-86-10.3 ~~ \today}

\maketitle

\begin{abstract}

We recall the early theoretical speculations on the possible explosive uses of antimatter, from 1946 to the first production of antiprotons, at Berkeley in 1955, and until the first capture of cold antiprotons, at CERN on July 17--18, 1986, as well as the circumstances of the first presentation at a scientific conference of the correct physical processes leading to the ignition of a large scale thermonuclear explosion using less than a few micrograms of antimatter as trigger, at Madrid on June 30th -- July 4th, 1986. 

\end{abstract}

\centerline{\emph{\underline{Preliminary remark}}}
%---------------------------------------------------

\emph{This paper will be followed by} by {\bf Antimatter weapons (1986-2006): From the capture of the first antiprotons to the production of cold antimatter}, \emph{to appear in 2006.}

\section{Introduction}
%=====================
\label{int:0}

 At CERN (the European Laboratory for Particle Physics), on the evening of the 17 to the 18 of July 1986, antimatter was captured in an electromagnetic trap for the first time in history.   Due to the relatively precarious conditions of this first successful attempt, it was only possible to conserve the antiprotons for about ten minutes.   This was, nevertheless, much longer than the Americans Bill Kells of Fermilab and Gerald Gabrielse of the University of Washington had hoped for.

 When these researchers return to CERN for another attempt, an improved apparatus will permit them to literally `bottle' several tens or hundreds of antiprotons.   Ultimately, the perfection of this technique will allow them to carry home a substance infinitely more rare and difficult to obtain than a piece of the Moon.   They would thus be able to complete, in their own laboratory, a most important experiment for the theory of the unification of the fundamental physical forces, that of comparing, with a precision greater than one part per billion, the masses of the proton and antiproton.

 Some other American Scientists, this time coming from the Los Alamos military laboratory (where the atomic bomb was perfected during the Second World War), are also at work in Geneva.   In a few months time, using many more resources and more sophisticated equipment, they also expect to capture and bottle antiprotons, but in much greater quantities.

 They will, as the group from the University of Washington, strive to divulge the difference in mass between the proton and its antiparticle.   But, they will also attempt a number of complex manipulations such as, the production of antihydrogen, the injection of antiprotons into superfluid helium, the search for metastable states in ordinary matter, etc.   Various crucial experiments that should, in the near future, help to determine whether or not antimatter could become a new source of nuclear energy for civilian and military applications.   For the more delicate experiments, they could certainly bring their vintage 1987 or 1988 bottles of antimatter to Los Alamos.   There, up in the peaceful mountains of New Mexico, they could perfect nuclear weapons free of radioactive fallout, beam weapons projecting thermonuclear plasma jets, gamma- or X-ray lasers, or other still more secret weapons, all triggered by antimatter.

\section{A concept born in 1946...}
%==================================
\label{con:0}

 Paradoxically, as futuristic and revolutionary as these weapons may seem, the military importance of antimatter \cite{GRINE1984-}, provided it can be produced, is as old as the science-fiction that has been talking about it.   For instance, it is quite possible that Edward Teller, the father of the American H-bomb, already had ideas of eventual military applications when he published in 1947, with Enrico Fermi, an article treating the capture of negative particles heavier than electrons by matter \cite{2}.   It is just as significant to notice that since 1945, about half of Teller's non-classified publications and many articles published by Andrei Sakharov \cite{3}, the father of the Soviet H-bomb, are concerned in one way or another with antimatter.

   In fact, it is in February 1946, in the first volume of \emph{The Bulletin of The Atomic Scientists} --- the first professional journal dedicated to nuclear arms-control and disarmament --- that the first reference ever is made to antimatter weapons.  In an article entitled ``Russia and the atomic bomb,'' as a further example of the Soviet press tendency towards sensational presentations, one can read concerning the possible discovery of antiprotons in cosmic rays:
\begin{quote}
\emph{Prof.\ D.V.\ Skobeltzyn of the Lebedev Institute directed a series of investigations of cosmic rays on Mt.\ Elbrus in the Caucasus.  It was apparently this group, which, assisted by Prof.\ Kapitza\footnote{P.L.\ Kapitza, winner of the 1978 Nobel Prize in Physics} in the contribution of a large magnet, recently made the discovery in cosmic rays of a new elementary particle --- the negatively charge proton.  (This discovery, too, was sensationalized in newspapers as a ``new way to produce atomic bombs!'')} \cite[p.~10]{BAS---1946-}.
\end{quote}
However, Western scientists doubted \cite{WANG-1946-}, and it finally turned out that in a letter followed by an article in the Physical Review\footnote{This was  before the Cold War, when East-West scientific exchange was relatively easy.} \cite{SKOBE1946-}, Skobeltzyn did not mention this experiments at all.  Indeed, detecting antiprotons in cosmic ray experiments \cite{VALLA1946-}, or in the fission of nuclei \cite{BRODA1946-}, is very difficult --- if not hopeless.  Nevertheless, as is demonstrated by the considerable interest aroused by these experiments, as well as by the number of published papers related to antiprotons in the immediate post-war period  \cite{HEILB1989-}, the year 1946 was an important date in the quest for antimatter.

  It is therefore quite clear, to give just one example, that a scientist like Teller who moved in February 1946 form Los Alamos to the University of Chicago (where \emph{The Bulletin of The Atomic Scientists} was edited and published) must have been fully aware that antimatter could open an alternate route to nuclear explosives \cite[p.174]{HEILB1989-}.  It could even be that this option was mentioned as a remote but worth investigating  possibility at the April 1946 secret Los Alamos meeting on the feasibility of the ``Super,'' i.e., the hydrogen bomb.

  As a matter of fact, in 1950, two years before the explosion of the first H-bomb, the ignition by antimatter of a mixture of deuterium and tritium was already being studied.   However, as shown for example in an article by A.S. Wightman \cite{4} (studying specifically the problem of the capture of antiprotons by deuterium and tritium), or in an article by J. Ashkin, T. Auerbach and R. Marschak \cite{5} (trying to calculate the result of the interaction between an antiproton and a nucleus of ordinary matter), the major problem at that time was that there wasn't any experimental data enabling to make a precise prediction of what would happen, for example, when a proton and antiproton met.   Nevertheless, well founded theoretical arguments already permitted a good understanding of the two essential characteristics of such a so-called annihilation reaction, a reaction in which the masses of a particle and its antiparticle are totally transformed into energy.

 These two characteristics are still valid today and entirely justify the interest in antimatter.   The first, is that the release of usable energy per unit mass is greater in annihilation than in any other nuclear reaction. One proton-antiproton annihilation releases 300 times more energy than a fission or fusion reaction.   The second, is that when antimatter is brought in the proximity of matter, annihilation starts by itself, without the need of a critical mass as in fission, and without the ignition energy needed in fusion.

 In short, an ideal nuclear trigger, provided methods to produce and manipulate sufficient quantities of antimatter be found.   But, at that time, the how and when antimatter could be produced wasn't known, and a number of fundamental questions about annihilation were still outstanding.   Consequently, for several years, applied research concentrated on more promising near term techniques, though less elegant for the theoreticians.   Thus the problem of igniting the H-bomb was resolved by using an A-bomb as a trigger, and the existence of the antiproton remained theoretical until 1955.

\section{The production of the first antiprotons}
%================================================
\label{pro:0}

 Historically, the first antiparticle ever observed was the antielectron, also called positron.   It was discovered in 1932 by Carl David Anderson, who while observing cosmic radiation, noticed a particle of the same mass as the electron, but of opposite charge.   Evidently many attempts were made to discover the antiproton, using the same method, but without success.   With the detectors available at that time and knowing only its mass and electrical charge, it was practically impossible to identify with any certitude the antiproton within the cosmic radiation.   It had to be artificially produced.   For that an accelerator, much more powerful than anything built up until that time, was needed.   Briefly, this is how antimatter is produced: protons are accelerated close to the speed of light, and then projected at a target.   The ensuing collision is so violent, that part of the energy is transformed into particle-antiparticle pairs.   Once this accelerator was built in 1955 at Berkeley, antiprotons were ``seen'' for the first time.

 By injecting them into a liquid hydrogen filled detector, the energy liberated in the explosive encounter of an antiproton and a proton, was seen to rematerialize into a scatter of other particles, essentially pions, shooting off in all directions, and carrying away with them most of the annihilation energy.

 But Edward Teller and his student Hans-Peter Duerr didn't stop there \cite{6}.   In 1956, they forwarded a hypothesis: If instead of annihilating with a simple hydrogen nucleus, the antiproton annihilated with a proton or neutron situated in the heart of a complex atom, such as carbon or uranium, the nucleus in question would literally explode.   This would result in a very large local energy deposition, thus bringing to light again, in theory, many civilian and military applications of antimatter.\footnote{In a 1986 telephone conversation with Andre Gsponer, Prof.~Hans-Peter Duerr at Max Planck Institute Munich said: ``Now I understand why Teller was so much interested in antimatter!''}

  At the experimental level, however, the Duerr-Teller hypothesis could not easily be tested: It required relatively low-energy antiprotons, and measuring techniques not yet available.  Research therefore turned towards substitute particles, such as negative kaons ($K^-$), which are short-lived but negatively charged and strongly interacting like antiprotons, and which moreover have a property called ``strangeness'' that enables to follow them through complicated interactions.  Since the study of these exotic and short-lived particles was primarily done at open laboratories dedicated to pure academic research, the reports documenting the military interest in their nuclear properties were confined to highly classified documents, e.g., pages 3 and 4 of Ref.~\cite{UCRL-1956-}.  It is only years later that explicit reports were published \cite{BLOOM1969-}

\begin{figure}
\begin{center}
\resizebox{8cm}{!}{ \includegraphics{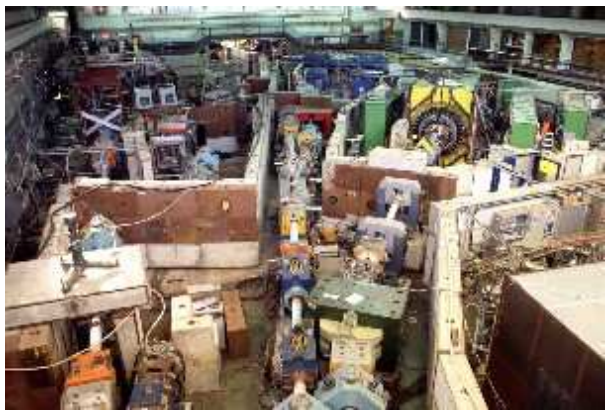}}
%..............................................
\end{center}
\caption[LEAR experimental area]{\emph{Overview of the LEAR experimental area. The Low Energy Antiproton Ring (LEAR) at CERN, is the only one of its kind in the world.   It enables scientists to study the details of antiproton interactions with the nucleus of atoms.   In the picture we can see the 80 meter in circumference ring that permits the storage and slowing of antiprotons down to energies as low as 5 MeV.   It's the first machine ever built to decelerate, rather than accelerate, particles. (Photograph: CERN Bulletin 47/98, 16 November 1998.)}} 
\end{figure}

 Thirty years passed by before the complex of machines necessary to accumulate and slow down antiprotons was conceived.   The only system of this type in the world \cite{7} is at CERN (Fig.1).   Finally, it was possible to study, on a large scale, the meeting of low-energy antiprotons with nuclei.   As a result, it has been possible to demonstrate that the energy deposition, although less than Teller (or others more recently \cite{8}) had hoped for, is sufficient to assure the feasibility of military applications of antimatter.   On the other hand, due to its very high cost and the enormous amount of energy needed to produce it, it has also become clear that antimatter could never become a usable source of energy for a power-plant.

 Thanks to the results of CERN, we were able to publish in August 1985, an estimation of the number of antiprotons needed to start thermonuclear reactions, be it to ignite an H-bomb or to trigger the microexplosion of a thermonuclear fuel pellet \cite{9}.   We thus discovered that it is possible to build a H-bomb, or a neutron bomb, in which the three to five kilograms of plutonium are replaced by one microgram of antihydrogen.   The result would be a bomb so-called ``clean'' by the militaries, i.e., a weapon practically free of radioactive fallout, because of the absence of fissile materials (Fig.2).

\begin{figure}
\begin{center}
\resizebox{8cm}{!}{ \includegraphics{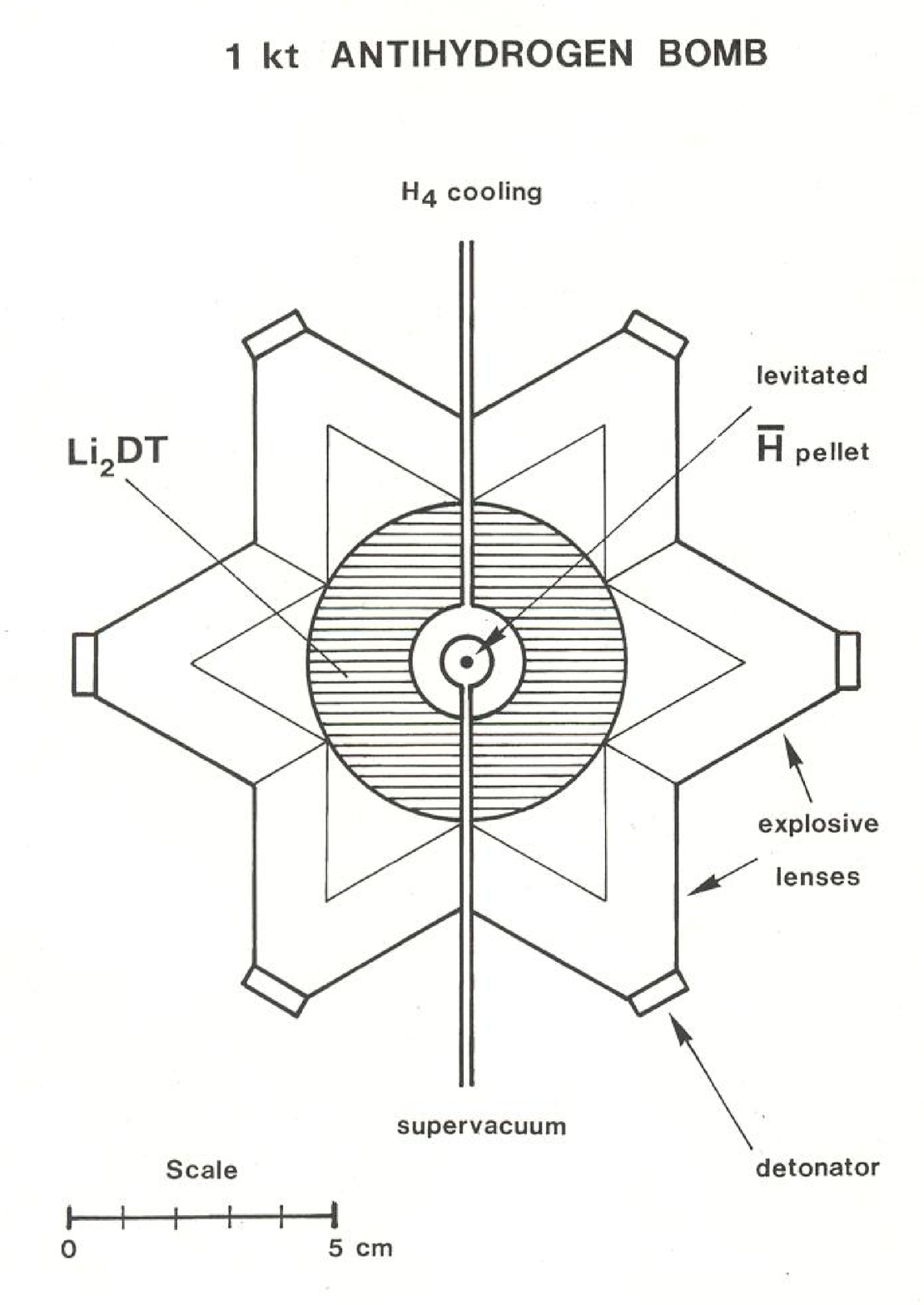}}
%..............................................
\end{center}
\caption[Antimatter bomb]{\emph{It is possible to construct a thermonuclear weapon in which the three to four kilograms of plutonium, necessary for the ignition, are replaced by one microgram of antihydrogen.   In this hypothetical bomb, the antimatter is in the center in the form of a pellet a tenth of a mm in diameter.   It is surrounded by, and isolated from, the thermonuclear fuel (a 100~g hollow sphere of $Li_2DT$).   After compression by explosive lenses, the fuel comes into contact with the antihydrogen.   Annihilation reactions start spontaneously, providing the energy to ignite the thermonuclear fuel.   If the chosen degree of compression is high, a bomb with increased mechanical effects is obtained, and if it is low, a neutron bomb.   In both cases the electromagnetic pulse effect and the radioactive fallout are substantially lower than that of a conventional A- or H-bomb of the same yield (1~kt).}} 
\end{figure}

\section{The revived military interest}
%======================================
\label{rev:0}

 For such a military use to be realistic, a technology capable of producing enough antiprotons for at least one antimatter trigger per day is needed.   This corresponds to a minimum production rate of $10^{13}$ antiprotons per second, six orders of magnitude higher than that at CERN today ($10^7$ antiprotons per second).   But, in theory, there exist numerous ways to increase this rate \cite{9}.   What we were unaware of, was that since the summer of 1983, the RAND Corporation had been carrying out a study for the U.S.   Air Force, ``examining the possibilities for exploiting the high energy release from matter-antimatter annihilation'' \cite{10}.   Similar concerns had equally sprouted-up in the Soviet Union \cite{11}.   The RAND study was completed in 1984.   The version published in 1985 constitutes a serious evaluation of the development possibilities of such an undertaking, in view of military applications.

 According to this document, a definitive evaluation of the possibility to produce and manipulate $10^{13}$ antiprotons per second, and the construction of transportable antiproton reservoirs, should be realized within the next five to seven years; many important technological problems being able to be studied with ordinary particles instead of antiprotons.   This same report mentions four main categories of applications: `propulsion' (fuel for ultra-fast anti-missile rockets), `power generators' (light and ultra-compact for military platforms in orbit), `directed energy weapons' (antihydrogen beams or pumped lasers relying on very short duration energy release) and `"classified additional special weapons roles"' (various bombs triggered by antimatter).

 In addition to the advantages related to its extremely high energy density and ease of ignition, annihilation has two important characteristics: the release of energy in a matter-antimatter explosion is extremely fast (ten to a thousand times shorter than a nuclear explosion), and most of the energy is emitted in the form of very energetic light charged particles (the energy to mass ratio of the pions emitted in annihilation is two thousand times higher than the corresponding ratio for the fission or fusion reaction products).   With the help of magnetic fields, very intense pion beams can be created, of the order of 100 mega-amperes per microgram of antiprotons.   Such beams, if directed along the axis of an adequate device, can drive a magneto-hydrodynamic generator, generate a beam of electromagnetic waves, trigger a cylindrical thermonuclear explosion, or pump a powerful X-ray laser.   In the last case, for example, the pions's energy could be used to transform in a very uniform plasma, a long cylinder of a substance such as selenium, whose ionized atoms have excited states favorable to the spontaneous emission and amplification of coherent X-rays.   But this is only one of the many concepts that permit, thanks to antimatter, to conceive X-ray lasers having efficiencies ten to a thousand times higher than those pumped by any other known energy sources.

 A certain number of experiments, that can only be carried-out with antimatter, are necessary to perfect these applications.   As long as antiprotons made in Europe (on Swiss Territory), could be bottled and brought back to the United States, the RAND Corporation concludes that a production/accumulation facility, such as the one at CERN, although desirable, wouldn't in the near future have to be built in the United States \cite[p.~43]{10}.

\section{Fundamental research or military research ?}
%====================================================
\label{fun:0}

 In view of its considerable strategic potential (for instance, antimatter seems to be a particularly interesting pump source for the Star War's X-ray lasers), it's not at all surprising that Soviet and American Scientists interested by the eventual applications of antimatter are eager to come to CERN, which at present has at least a five year lead in antimatter technology.   In this context, it also wouldn't be surprising if a blunder was made...

 In effect, for the teams of American physicists coming from weapons laboratories, the official justification for their coming to CERN, is to carry-out fundamental research, pure scientific research.   In the beginning of July 1986, these same Americans were supposed to go to Madrid, where a full session of the Fourth International Conference on Emerging Nuclear Systems was dedicated to antimatter energy concepts.   At this same conference we were to present the point of view that the only realistic applications for annihilation energy were in the military domain \cite{13}.

 To everyone's surprise, the Americans didn't come.   Ten days before the conference, they announced their withdrawal without giving any convincing explanation.   The participants quickly realized that the American authorities had undoubtly reevaluated the military importance of antimatter, and had probably prevented the Los Alamos Scientists from coming to Madrid \cite{14}.   Thus exposing that scientists working at CERN, and coming from a non-European weapons laboratory, had other than fundamental research interests, that were obviously militarily sensitive.

\section{Strategic and political consequences}
%=============================================
\label{str:0}

 Whether antimatter triggered thermonuclear weapons are realizable or not, or whether other weapons using annihilation energy are feasible or not, the fact that a relatively small quantity of antimatter can set off a very powerful thermonuclear explosion creates serious problems for the future of the strategic balance \cite{BROAD1985-}.   In fact, the arms control treaties presently in force deal only with fission related devices and materials \cite{15}: atomic bombs, nuclear reactors and fissile materials.   By removing the fission fuse from thermonuclear weapons, antimatter triggered H-bombs and neutron bombs could be constructed freely by any country possessing the capacity, and be placed anywhere, including outer-space.

 Then again, even if technical obstacles prevented, for example, the actual construction of battle-field antimatter weapons, antimatter triggered microexplosions would still allow small and middle sized thermonuclear explosions to be made in the laboratory.   This possibility would considerably reduce the need for underground nuclear explosions, thus rendering ineffective any attempt to slow the arms race by an eventual comprehensive nuclear test-ban treaty \cite{15}.   A nuclear test laboratory of this type could be based around a large heavy-ion accelerator \cite{16}, which would provide a means of massive antimatter production, as well as a driver to study the compression and explosion of thermonuclear fuel pellets.

\section{CERN convention and CERN management}
%============================================
\label{CERN:1}

Instead of a conclusion, or an appraisal of the original version of this paper (which was published in 1986, see Ref.~\cite{GSPON1986-}) we quote in this final section point~1 of article~II of the Convention defining the main purpose of CERN, and two examples of how the letter of this article was interpreted by members of the CERN mangement in October 1985 at the colloquium celebrating the 40th anniversary of the French Atomic Energy Commission \cite{CEA--1986-}, nine months before a team led by a US scientist sponsored by the US Air Force, captured antimatter for the first time in an electromagnetic trap.

\subsection{CERN convention, Article II: Main purpose.}
%......................................................

\begin{quote}
\emph{ARTICLE II : Purposes. 1. The Organization shall provide for collaboration among European States in nuclear research of a pure scientific and fundamental character, and in research essentially related thereto. The Organization shall have no concern with work for military requirements and the results of its experimental and theoretical work shall be published or otherwise made generally available} \cite{CERN-1953-}. 
\end{quote}

\subsection{CERN Director general and Nobel laureate: Carlo Rubbia.}
%...................................................................

\begin{quote}
Carlo Rubbia:\footnote{In 1985, Carlo Rubbia was foreseen to become the next Director general of CERN.}~
\emph{``Antimatter is (...) produced at CERN where a real production factory has been built.  We accelerate protons and let them strike a target.  In these collisions many particles and antiparticles are produced, and, among them, one finds antiprotons.  These antiprotons are stored in a magnetic bottle which, in our case, is a storage ring. On obtains this way a quantity of antimatter that is not very small as it amounts to one nanogram\footnote{Translator's note: Not microgram, i.e., 1000 nanograms, as in the French text.} per day. But I would like to speculate on all what could be done if one had larger quantities of antimatter, say on the order of a few grams. One gram of antimatter annihilating with matter can produce an energy equivalent to that obtained in burning 10'000 tons of hydrogen-oxygen fuel.  One has therefore in one gram the energy content of 10'000 tons of a high efficiency fuel...\\
~\\
I think that this property of antimatter --- storage of enormous amounts of energy in a very small volume --- can be very interesting, for instance in astronautics.  In effect, to send something from Earth into outer-space, one has to consume about hundred times the weight of the payload in the form of fuel.  However, if one uses antimatter as the energy source, one could send objects without being penalized at the payload level. I have calculated that with one milligram of antimatter one could envisaged an Earth-Mars round trip.  It is therefore clear that absolutely fantastic possibilities are offered to us for all applications in which a large concentration of energy is useful.\\
~\\
Indeed, it is not excluded --- and this is even possible at present --- to build machines able to accumulate much larger quantities of antimatter than at the moment.  We are at CERN very inefficient in collecting antimatter, but one can envisage other techniques allowing a thousand-fold increase in the accumulation efficiency. For example, one could store this antimatter in the form of an antiproton-positron plasma in the same type of magnetic bottles than those used in fusion\footnote{This is a seldom mentioned nuclear weapons proliferation implication of thermonuclear fusion reactors \cite{GSPON2004A}.} and this stored antimatter would be available for practical use.\\
~\\
Antimatter opens a whole range of applications and these ideas are worth being studied''} \cite[p.~120-122]{CEA--1986-}.
\end{quote}

\subsection{CERN Director of research: Robert Klapisch.}
%...................................................................

\begin{quote}
Robert Klapisch:~
\emph{I would like to further discuss the certainly futuristic proposals made by Carlo Rubbia a short while ago concerning, in particular, antiprotons.  
One has to remember that the antiproton has been discovered about 30 years ago by a research team at Berkeley: Segre, Chamberlain, ... At the time there were probably a dozen antiprotons, subject of this discovery. I do not know how long this experiment took, but assume that it took one month:  Well, you can see that in 30 years one has progressed by eleven orders of magnitude in the capacity of producing and storing antimatter.  What Carlo Rubbia is proposing, in dreaming a little bit, is in a way to go further by twelve orders of magnitude, and one cannot exclude --- because the physical principles are known --- that in two or three decades one reaches this goal.\\
~\\
But closer to us, I mean 3--4 years from now, CERN could, with facilities in existence or in project, supply antiprotons in a bottle, a kind of a large cryogenic and magnetic Dewar.   Not one gram, but $10^8$ or $10^9$ antiprotons that could be put on a truck and transported to another research laboratory.  The real problem is to know what one will do with that amount of antimatter, as much in the perspective of science than of applications.  There, I would like to recall a phrase that is attributed to one of the discoverers of the laser, who in 1960 said that the laser is a solution looking for a problem. I believe that this is the true question with antimatter; effectively there are technical possibilities; I am intuitively convinced that there are possibilities.  The real question now, is to see to what they may apply''} \cite[p.~129]{CEA--1986-}.
\end{quote}

\section{Appendix: Production and storage of antiprotons}
%========================================================
\label{app:0}

Relativistic quantum theory predicts the existence of two types of elementary particles appearing on an equal footing with respect to the fundamental equations.   Thus, for each particle there exists an antiparticle having the same mass and spin but opposite electrical charge.   Furthermore, particles and antiparticles can appear or disappear in pairs, due to the transformation of energy into matter and vice-versa.

 Antiprotons and positrons are probably the only forms of antimatter that will be able to be fabricated, in substantial quantities, in the near future.   They are produced by accelerating protons (or other particles) to energies such that, when they collide with a target, a part of the energy is transformed into particle-antiparticle pairs.   In practice, when using a fixed target, as a function of invested energy, the maximum antiproton production yield occurs when the protons are accelerated to an energy of about 120 GeV \cite{9}.   Since less than one collision out of thirty produces an antiproton, and since the mass of an antiproton corresponds to only 0.94 GeV, the energy efficiency is very poor.   From this point of view, a better solution would be to use a collider-ring in which the antiprotons would be produced by the head-on collisions of protons turning in opposite directions \cite{A}.   In theory, an even higher yield could be obtained if conditions similar to the original ``Big Bang'' could be recreated in the laboratory, conditions in which proton-antiproton production becomes spontaneous, a possibility that was first discussed by Edward Teller \emph{et al.}~\cite{B}.   Such conditions might be found in quark-gluon plasmas, which could be produced in high-energy heavy-ion collisions, which are presently the subject of intense research \cite{C}.

 Once the antiprotons are created (with a whole spectrum of velocities and directions), the following step consists of capturing them before they interact with matter.   This is a problem much more difficult to resolve than that of production.   It took almost thirty years before a solution was found at CERN.   This required the invention of ``stochastic cooling,'' a technique to decrease the width of the antiproton velocity distribution \cite{D}.   It is then possible to concentrate the collected antiprotons into a very small beam, to accumulate them in storage rings, and finally slow them down to energies such that they can be brought to a standstill in electromagnetic traps.

\begin{figure}
\begin{center}
\resizebox{8cm}{!}{ \includegraphics{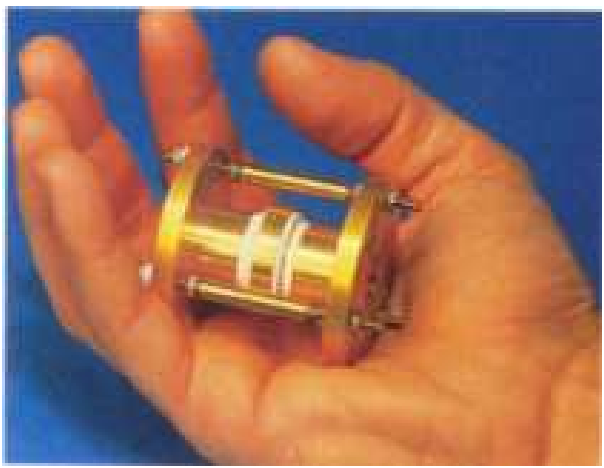}}
%..............................................
\end{center}
\caption[Antimatter bomb]{\emph{Small antimatter trap. (Dan Brown, Angels and Demons, Atria Books, 2005, ISBN 0-7432-7506-3, p.~71. Credit: CERN/ Photo Researcher, Inc.)}} 
\end{figure}

 In a Penning trap, particles are radially confined by a magnetic field, and axially by an electrostatic field.   A cylindrical trap of this type served as host during the recent experiments at CERN in which antiprotons were bottled for the first time.   It also trapped continuously a single electron for more than ten months at the University of Washington \cite{E}.   To store antiprotons for years, one needs a vacuum better than $10^{-18}$ torr.   This is obtainable only in enclosures that are sealed (after filling) and cooled to the temperature of liquid helium.   It is therefore practically impossible to measure the vacuum level, so that doing the experiment itself is the only way to verify the technique.   If this method is successful, it will be possible to make transportable bottles with a capacity of $10^{12}$ to $10^{13}$ antiprotons \cite{F}.

 Then the decisive stage for the practical applications of antimatter will begin: will it be possible to develop adequate simple and compact storage techniques? For this, two major approaches are being considered.   The first consists of making antihydrogen by combining antiprotons with positrons, and then trying to form solid antihydrogen pellets which could be stored and manipulated with the help of various electromagnetic and optical levitation techniques.   Very high storage densities would be obtained, but only in cryogenic enclosures and extremely good vacuums.

 The most appealing approach would be to store the antiprotons in ordinary matter.   In fact, if all antimatter particles have a tendency to spontaneously annihilate when coming into contact with matter (be it the effects of electromagnetic attraction in the case of positrons and antiprotons, or van der Waals forces for antihydrogen), the existence of metastable states of antiprotons in condensed matter can not be ruled out a priori \cite{G}.   For example, if a very low energy antihydrogen atom is diffused into a solid, it moves about until its positron annihilates with an electron.   The antiproton may then take the place of this electron, and under some conditions, remain confined at certain points within the crystalline structure.   At present the kind of substance to be used isn't known, but an enormous variety of chemical compounds and crystal types are available for the search of an optimum material.

 Other less obvious solutions could still be discovered.   For example, antiprotons might, as electrons do when placed in liquid helium, form a bubble at the center of which they could subsist indefinitely \cite{G}.   Also, similar to the electron pairs responsible for superconductivity, antiprotons might possibly form Cooper pairs if placed in a metal, becoming thereby unable to lose kinetic energy by shock, and thus to annihilate.

\end{document}